# Quantum mechanical analysis of nonlinear optical response of interacting graphene nanoflakes

Hanying Deng,[1,2] David Zs. Manrique,[2] Xianfeng Chen,[1] Nicolae C. Panoiu,[2] and Fangwei Ye[1,a]

[1]Key Laboratory for Laser Plasma (Ministry of Education), Collaborative Innovation Center of IFSA (CICIFSA), School of Physics and Astronomy, Shanghai Jiao Tong University, 800 Dongchuan Road, Shanghai 200240, China

[2]Department of Electronic and Electrical Engineering, University College London, Torrington Place, London WC1E7JE, United Kingdom

[a]Corresponding author: Fangweiye@sjtu.edu.cn

## Abstract

We propose a distant-neighbor quantum-mechanical (DNQM) approach to study the linear and nonlinear optical properties of graphene nanoflakes (GNFs). In contrast to the widely used tight-binding description of the electronic states that considers only the nearest-neighbor coupling between the atoms, our approach is more accurate and general, as it captures the electron-core interactions between all atoms in the structure. Therefore, as we demonstrate, the DNQM approach enables the investigation of the optical coupling between two closely separated, but chemically unbound GNFs. We also find that the optical response of GNFs depends crucially on their shape, size, and symmetry properties. Specifically, increasing the size of nanoflakes is found to shift their accommodated quantum plasmon oscillations to lower frequency. Importantly, we show that by embedding a cavity into GNFs, one can change their symmetry properties, tune their optical properties, or enable otherwise forbidden second-harmonic generation processes.

# I. INTRODUCTION

Owing to their ability to confine and guide light down to nanometer scale, the collective oscillations of electrons in conducting materials, known as plasmons, have generated great expectations for applications ranging from metamaterials[1–3], quantum optics[4,5], and photovoltaics[6] to photodetectors[7] and biological sensing[8,9]. Plasmons are commonly observed in noble metallic nanostructures, appearing as pronounced spectral resonances in their optical absorption and scattering spectra. However, noble metals suffer from a relatively large ohmic losses, resulting in a limited plasmon lifetime[10, 11]. In addition, metal plasmons can hardly be tuned, unless one uses metallic nanoparticles with different shapes[12], thus severely limiting the operating spectral domain.

Graphene, a monolayer of carbon atoms arranged in a hexagonal lattice[13], has emerged as a promising alternative to noble metals for nanoplasmonic applications due to its ability to support plasmons with unique properties, including large tunability[14, 15], long lifetime[15, 16], and high degree of optical confinement[17, 18]. Strong intrinsic optical nonlinearity has also been observed in graphene and that can be further enhanced by plasmons. By using bottom-up chemical synthesis methods[19] or top-down electron-beam techniques[20, 21], nanometer-sized graphene nanoflakes (GNFs) with various shapes and sizes can be manufactured, providing a versatile platform to investigate plasmonic phenomena at quantum level. In addition to inheriting remarkable physical properties from the extended sheet, the nanometer-sized graphene supports even more confined plasmons. Moreover, the plasmonic response of nanometer-sized graphene can reach the visible-light spectrum region that is beyond the range of the plasmons in extended graphene sheets, thus extending our ability to manipulate visible light at deep-subwavelength scale.

Intense research efforts have recently advanced our understanding of linear and nonlinear optical properties of nanostructured graphene[22–25]. For extended graphene, the nonlinear response is commonly described using a classical nonlinear conductivity derived from the Boltzmann transport equation[26], assuming that intraband transitions dominate the optical response. Such classical electrodynamic description is invalid for nanometer-sized graphene, as the optical properties of nanostructured graphene are strongly influenced by nonlocal and finite-size effects. More recently, *ab inito* methods and tight-binding (TB) description of the electronic states have been applied to investigate the optical response of the nanometer-size graphene[27, 28]. The TB method is not accurate enough in general, since only the interaction of nearest-neighbor atoms is considered. Moreover, the TB method is not suitable for non-tightly bound structures, such as two interacting GNFs. Although *ab inito* techniques take into account the many-body interactions and therefore are more accurate and general than the TB approach, quantitatively accurate predictions of optical properties are computationally expensive and therefore their application are limited to fewer atoms than those employing the TB method.

In our study, we use an all-atom coupling approach to calculate the linear and nonlinear polarizabilities of GNFs, which are the finite-size analog of linear and nonlinear optical susceptibilities of bulk optical media. In our approach, we assume that the $\pi$-orbitals of each atom are coupled to the core potential of all atoms. The electronic structure is then calculated and a perturbative approach[29] is subsequently used to evaluate the linear and nonlinear polarizabilities. Importantly, our method inherently accounts for the symmetry properties of GNFs. For example, our method predicts that no second-harmonic is generated in GNF configurations invariant to inversion symmetry transformations as SHG is forbidden in such centrosymmetric structures. In order to illustrate the flexibility and generality of our method, we apply it to GNFs of different shape and size, cavities in GNFs, and dimers of GNFs of different shapes.

## II. Optical response calculation and Results

### A. Optical response calculation

Under illumination with an optical beam with frequency tuned to a frequency at which plasmons can be excited in the GNF, these graphene structures generate significant nonlinear optical response, which manifests as enhanced second-harmonic generation (SHG) and third-harmonic generation (THG). In this work, we study the nonlinear optical response of GNF with various shapes and topologies, namely the linear and nonlinear polarizabilities of such structures. We model GNFs as a planar hexagonal distribution of carbon atoms with lattice constant $a = 1.42$ Å. For example, as illustrated in Fig. 1(a), a triangular GNF consisting of 141 carbon atoms has a side length of 2.2 nm. Considering that the localized electrons in the carbon-carbon $\sigma$-bond do not contribute significantly to the low energy optical response, only the $\pi$-bond forming $p_z$ orbitals are included in the quantum-mechanical calculation. That is, each carbon atom is represented by a single $p_z$ orbital oriented perpendicularly to the graphene plane, as per Fig. 1(b).

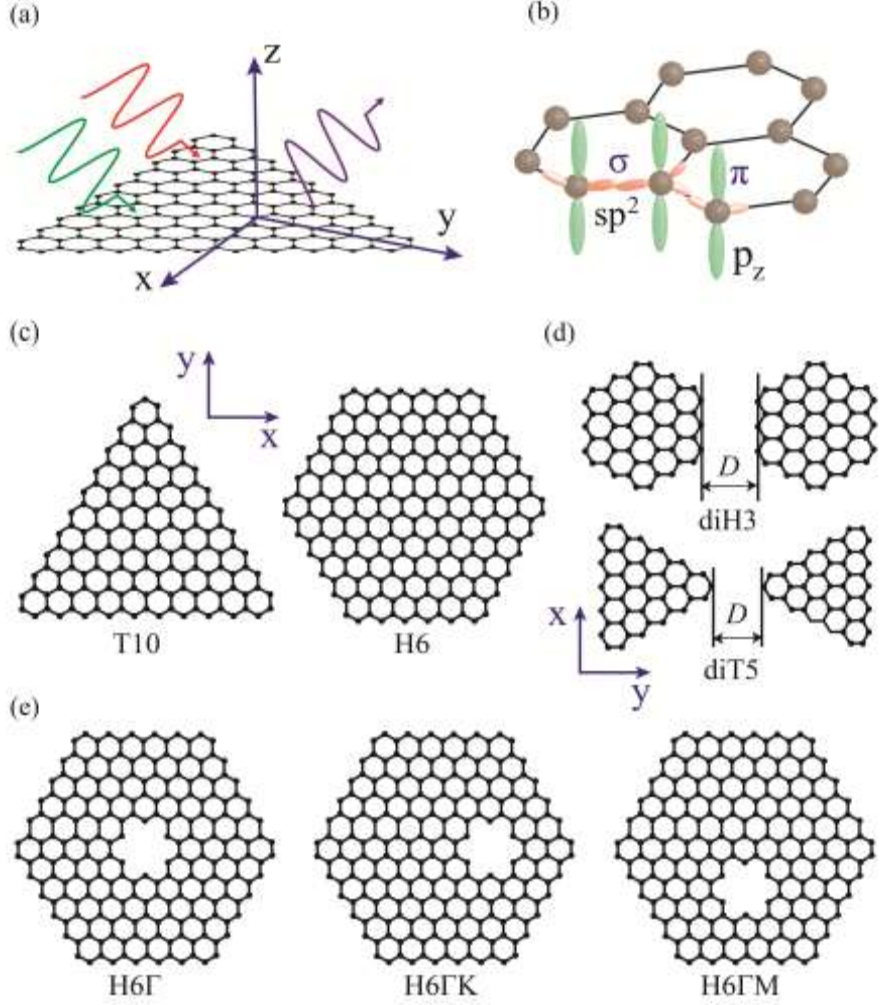

**FIG. 1.** (a) Illustration of nonlinear optical interactions in a triangular GNF. (b) Schematic representation of $sp^2$ hybridization of carbon atoms of graphene. (c) Triangular and hexagonal GNFs. T$n$ and H$n$ denote triangular and hexagonal GNFs, where $n$ is the number of 6-atom hexagons on the side of GNFs. (d) Dimer structures with separation distance, $D$. (e) Hexagonal GNFs containing a cavity at the $\Gamma$ point and along the $\Gamma M$ and $\Gamma K$ directions. In all cases, the planar GNFs lie on the $x-y$ plane.

In order to investigate non-tightly bound structures, we considered interactions between the $p_z$ electrons and all cores in the structure. The Hamiltonian operator for a single electron in the GNF is expressed as:

$$\hat{H} = -\frac{\hbar}{2m}\nabla^2_{\vec{r}} - \sum_{\alpha=1}^{N} \frac{Z_{eff}e^2}{\vec{r}-\vec{r}_{0\alpha}}, \qquad (1)$$

where $Z_{eff}$ is effective core charge, the first term describes the kinetic energy, and the second term represents the potential arising from Coulomb interactions between the $p_z$ electron and all the other nuclei in the GNF. The electronic structure is computed using the Schrödinger equation

$$\hat{H}\psi(\vec{r}) = E\psi(\vec{r}), \tag{2}$$

where $E$ and $\psi(\vec{r})$ are the eigenenergy and eigenfunction, respectively. Then, we follow a perturbative approach to evaluate the linear and nonlinear optical response of GNFs (see supplementary material for more details about the quantum mechanical calculations). The structures investigated in this paper are a series of triangular and hexagonal GNFs (Fig. 1(c)), and dimers of triangular and hexagonal GNFs (Fig. 1(d)) to demonstrate inter-nanoflake optical coupling. We also study the optical response of hexagonal GNFs with an embedded cavity, as shown in Fig. 1(e). The cavity is located either at the center of the cavity or along the $\Gamma M$ or $\Gamma K$ symmetry axes of the cavity.

**B. Shape and size effects in optical response of graphene nanoflakes.**

We assume that the incident electric field is polarized in a direction parallel to one of the sides of the triangle or hexagon, indicated as the $x$-axis in Fig. 1(a), with all the structures considered in this work being assumed to lie in the $x−y$ plane. We first consider hexagonal GNFs. Because they are centrosymmetric structures, SHG is forbidden in hexagonal GNFs and thus we focus on their first- (linear) and third-order optical response.

In Fig. 2, we show the frequency dependence of linear polarizability and THG nonlinear polarizability of hexagonal GNFs, with the size of the GNF increasing from the top to the bottom panel, as illustrated in the middle panels. Figures 2(a)-2(f) present the real and imaginary part of linear polarizability, $\alpha_{xx}(\omega)$, of hexagonal GNFs. The polarizability spectrum clearly shows that the peak of the imaginary part of the polarizability occurs at the spectral position where its real part becomes exactly zero, which is the condition that defines the existence quantum plasmons. For the smallest size, which is a single Benzene, the linear polarizability shows a resonance at 6 eV (see Fig. 2(a)), which corresponds to the transition from the highest-occupied molecular orbital (HOMO) to the lowest-unoccupied molecular orbital (LUMO). With the structural size increasing, some additional resonances expectedly appear as more energy levels are involved into the optical transitions. As expected, our calculations predict that the magnitude of the polarizability increases when the size of the GNF increases, as the density of states increases with the number of atoms in the GNF. We also note that the peak energy of the plasmon resonance is continuously red-shifted with the increase in the structural size of the GNF, which is a consequence of finite-size effects of these GNF that is not included in their classical description but captured by our DNQM description. Importantly, as the size of the GNF increases additional plasmon resonances can be seen in the spectra, which is similar to what can be observed in the classical regime.

Figures 2(g)-2(l) show the real and imaginary part of third-order polarizability, $\gamma_{xxxx}(\omega)$, (also called second hyperpolarizability when referring to molecules) associated with the THG. Note that the nonlinear response of GNFs is also highly sensitive to the structural size, and with the increase of the size of the GNF the magnitude and the resonance frequency of the nonlinear plasmon oscillations exhibit the same trend as in the linear case. Also note that both linear and nonlinear resonance can reach visible-range spectrum as long as the size of the GNFs is small enough.

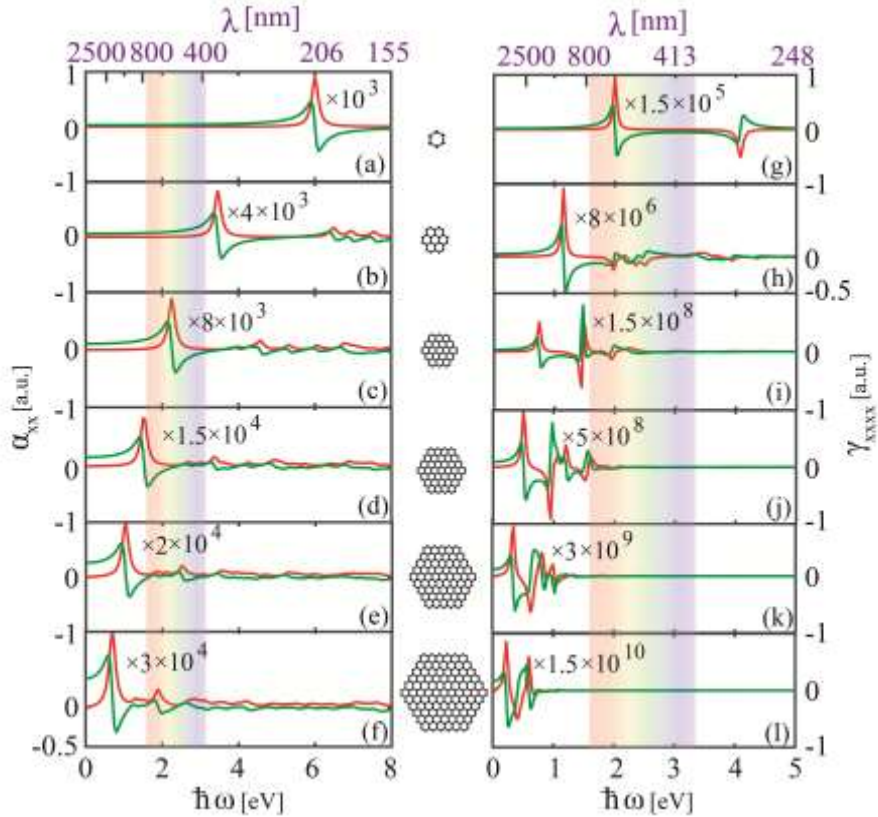

**FIG. 2.** The effect of GNF size on the optical response of hexagonal GNFs. Real and imaginary part of linear polarizability (a)–(f), and third-order nonlinear polarizability for THG (g)–(l) of hexagonal GNFs with different sizes as shown in the middle panel. The red (green) curves indicate the imaginary (real)

part of the polarizabilities. The linear and nonlinear polarizabilities are calculated in atomic units (a.u.) with $e = \hbar = m_e = a_0 = 1$, and should be multiplied by the factors given in the supplementary material.

The results for triangular GNFs are presented in Fig. 3. The top panels show the structures of triangular GNFs under consideration with their size increasing from left to right. Unlike the centrosymmetric hexagonal structures, triangular GNFs are non-centrosymmetric and consequently the second-order polalizability tensor, $\beta$, (also called hyperpolarizability) has non-zero components, meaning that the SHG is allowed. In particular, considering the symmetry properties of the triangle, $\beta_{yyy} \neq 0$, and thus this is the component we considered in our study.

The incident-frequency dependence of the linear polarizability $\alpha_{yy}(\omega)$ of triangular GNFs, calculated for GNFs of different size, is presented in Fig. 3(a). Note that increasing the structural size induces a red-shift for the plasmon frequency, similar to the case of hexagonal GNFs. The imaginary and real part of nonlinear polarizabilities corresponding to SHG ($\beta_{yyy}$), and THG ($\gamma_{yyyy}$) for triangular GNFs are shown in Figs. 3(b) and 3(c), respectively. Similarly to the case of hexagonal structures, the most pronounced peak in the spectra of nonlinear polarizabilities is red-shifted when the GNF size increases. However, the particular shape of the corresponding linear and nonlinear spectra of the triangular GNFs is markedly different from the analog ones calculated for the hexagonal GNFs, implying that the optical response of GNFs is strongly dependent on their geometrical configuration.

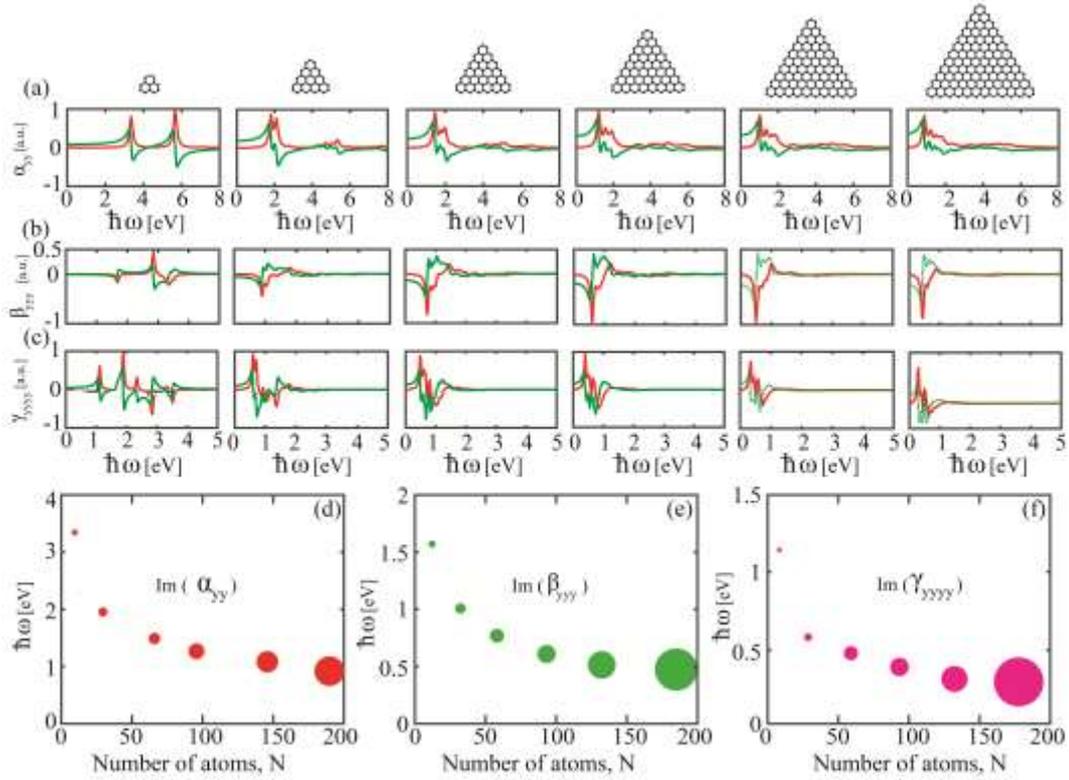

**FIG. 3.** The effect of GNF size on the optical response of triangular GNFs. We show the normalized real and imaginary part of (a) linear, (b) second-order, and (c) third-order polarizability, calculated for the triangular GNFs shown in the top panel. The red (green) curves indicate the imaginary (real) part of the polarizabilities. The circular symbols show the dependence of the first resonance peak energy on the number of carbon atoms for (d) linear, (e) second-order, and (**f**) third-order polarizabilities. The magnitudes of the peaks of the linear and nonlinear polarizabilities are proportional to the area of the circles.

In Figs 3(d), 3(e) and 3(f), we present an overview of the dependence of the first resonance peak energy for the first-, second-, and third-order polarizabilities, respectively, on the number of carbon atoms of a triangular GNF. Expectedly, the magnitude of the linear and nonlinear polarizabilities increases with the number of carbon atoms in the GNF. Additionally, the resonance peaks display an obvious red-shift with the increase of the number of carbon atoms, the steepest variation occurring in the linear response. In particular, when the number of carbon atoms increases from 13 to 198, the first resonance peak energy of linear, second-order, and third-order polarizabilities varies by 2.45 eV, 1.25 eV, and 0.825 eV, respectively.

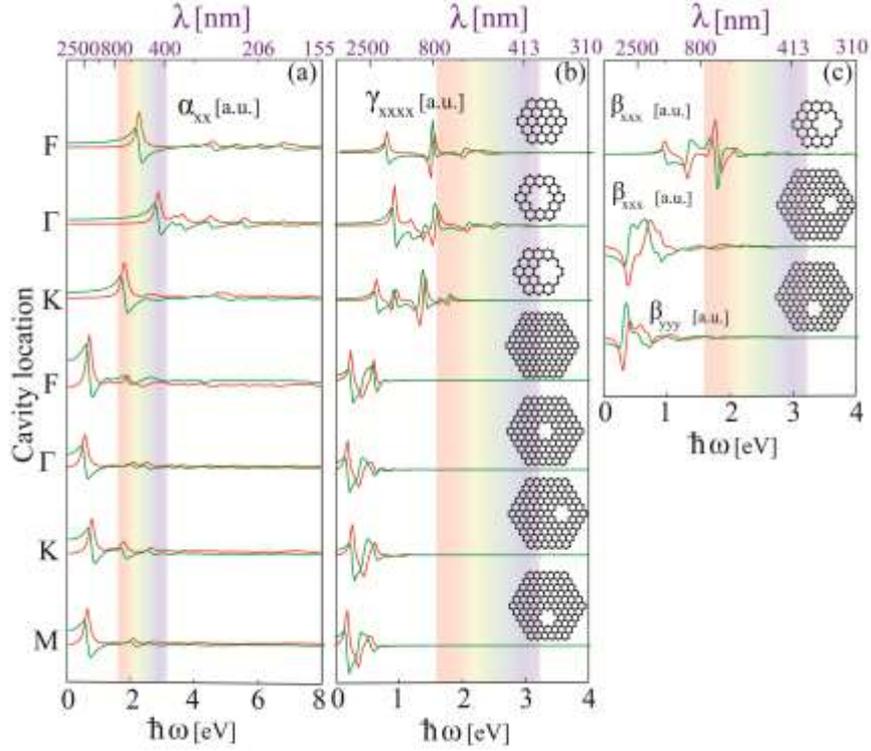

**FIG. 4.** Linear and nonlinear polarizabilities of hexagonal GNFs containing a quantum cavity. We consider the hexagonal GNF H3 with a cavity located at Γ and off center, along the ΓK direction, and H6 with a cavity at Γ or along ΓK and ΓM directions (see the middle panel). We show the normalized real (green) and imaginary (red) part of (a) linear and (b) the third-order polarizabilities. The normalized nonzero second-order polarizabilities when the cavity is along the ΓK and ΓM directions are shown in (c).

## C. Graphene nanoflakes containing a quantum cavity.

We next investigate the linear and nonlinear optical response of GNFs containing a quantum cavity. The cavity can be created at desired locations by removing one or more carbon atoms. As expected, the optical response of GNFs can be significantly altered by the presence of the cavity. In particular, the cavity can break the inversion symmetry of the GNF so that, as is the case with SHG, forbidden nonlinear optical interactions become possible. We illustrate this and other cavity effects in Fig. 4, by comparing the linear and nonlinear polarizabilities of two hexagonal GNFs (H3 and H6) that contain a cavity at different locations (see the cavity location in the inset). For a better comparison, the linear and nonlinear spectra of a full GNFs are also presented.

These calculations reveal several important conclusions. Thus, compared to the GNF of larger size (H6), the presence of a cavity into a smaller GNF (H3) produces more dramatic modifications of the spectra of polarizabilities. Interestingly, we note that a cavity introduced at the center of the H3 nanoflake induces a blue-shift of the linear and nonlinear plasmon frequencies, whereas a cavity located off center on the $\Gamma K$ symmetry axis induces a red-shifted (see the top panels in Figs. 4(a) and 4(b)). Equally important, we find that the H3 flake has resonances in the visible domain, even when it contains a cavity.

For the larger GNF (H6), the spectral changes produced when a cavity is inserted in the nanoflake are relatively small; however, in the presence of the cavity the second-order polarizability is no longer identical to zero. To be more specific, in the case of full hexagonal structures, SHG is strictly forbidden due to inversion symmetry, whereas in the presence of the cavity along the $\Gamma K$ or $\Gamma M$ symmetry axes, inversion symmetry breaking induces intense SHG when the fundamental frequency is near the plasmon resonances, as shown in Fig. 4(c). Of course, when the cavity is at the center of the GNF, the inversion symmetry of the structure is preserved and thus all even-order polarizabilities of hexagonal GNFs identically vanish. This effect can be used as an efficient approach to engineer the nonlinear optical response of graphene structures. It can also have important practical implications, e.g. to molecular sensing. Thus, similar to the effect of the cavity, a molecule adsorbed by the GNF can drastically alter its symmetry properties and significantly change the nature of generated nonlinear optical signal.

**D. Optical response of GNF Dimers.**

Finally, we explore the effects arising from optically coupling two closely spaced GNFs. As two specific examples, we investigate the optical coupling between two hexagonal and triangular GNFs, as schematically shown in Fig. 5. Here we want to emphasize that the widely used tight-binding model does not describe the electronic states for GNFs dimers, as it does not allow coupling between atoms beyond the nearest neighbors, and accordingly is not suitable for the study of the interaction between two GNFs. By contrast, the DNQM approach we proposed captures interactions among all atoms of the structure and thus it is well suited for the investigation of composite configurations, such as GNF dimers.

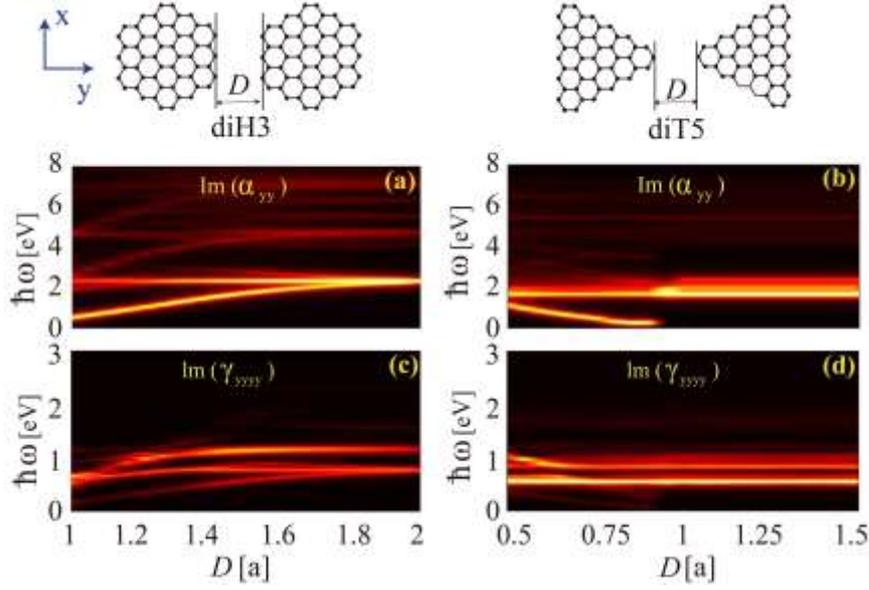

**FIG. 5.** Dependence of linear and nonlinear polarizabilities of GNF dimers on the separation distance, *D*. $a = 1.42$ Å is the lattice constant. (a), (b) Variation of the linear polarizibility of hexagonal and triangular GNF dimers with *D*, respectively. (c), (d) Variation of the third-order nonlinear polarizibility of hexagonal and triangular GNF dimers with *D*, respectively.

Figure 5 presents the variation of linear and third-order nonlinear polarizabilities of hexagonal and triangular GNF dimers as a function of the spacing between the two comprising nanoflakes, *D*. Note that the second-order nonlinear processes are forbidden in these two dimer configurations due to their inversion symmetry. As Fig. 5 shows, for large *D* values, the coupling between two flakes is negligible and consequently the spectrum of the "dimers" is essentially identical with that of a single GNF. Only when *D* becomes small enough the optical properties of a dimerized GNFs begin to deviate from those of a single GNF. Remarkably, as can be seen from the linear polarizability spectra shown in Figs. 5(a) and 5(b), whereas only one pronounced peak exits for a single-GNF as seen for larger *D*, for small *D*, the interaction between the two plasmons of the GNFs induces an energy split that increases as *D* decreases. In addition, with the increase of *D*, the additional lowest-energy peak is blue-shifted for the hexagonal GNF dimer (Fig. 5(a)) and is red-shifted for the triangular GNF dimer (Fig. 5(b)). The size of the energy splitting is an indication of the strength of the interaction between the GNFs. Using this measure, one can conclude that hexagonal GNFs interact much more strongly as compared to the triangular ones, a possible explanation of this finding being that more carbon atoms are in close proximity of each other than in the case of triangular GNF dimers. It should also be noted that the coupling between two GNFs can significantly alter the third-order nonlinear polarizabilities, too, as per Figs. 5(c) and 5(d).

## III. CONCLUSIONS

In summary, we have computed the linear and nonlinear polarizabilities of GNFs using a distant-neighbor quantum mechanical approach and revealed that the optical response of graphene nanoflakes depends significantly on their shape, size, and symmetry properties. In particular, the peak energy of the plasmon resonance is red-shifted and the magnitude of the linear and nonlinear polarizabilities increases as the size of the graphene nanoflake increases. Significant changes in the optical response were demonstrated by introducing a quantum cavity in the graphene nanoflake. In particular, strong second-harmonic generation is enabled in hexagonal graphene nanoflakes by using a cavity to break the structural inversion symmetry. We have also explored the optical response of graphene nanoflake dimers, and found that the strong coupling between two closely spaced graphene nanoflake leads to the energy splitting of the plasmon band, the magnitude of this splitting depending on the shape of the interacting nanoflakes. Importantly, the theoretical techniques and ideas developed in this study are not specific to graphene but can be extended to other two-dimensional (2D) materials, such as $MoS_2$ [30, 31] and black phosphorous[32, 33] (one only needs to update in our method the corresponding Hamiltonian operator and basis functions of each atom for these new materials). Potential applications of our findings to molecular sensors have been discussed, too.

## SUPPLEMENTARY MATERIAL

See supplementary material for the supporting content.

## ACKNOWLEDGMENTS

The work of H.D. and F.Y. was supported by Innovation Program of Shanghai Municipal Education Commission (Grant No. 13ZZ022) and the National Natural Science Foundation of China (Grants No. 61475101). N.C.P. acknowledges financial support from the European Research Council/ERC Grant Agreement No. ERC-2014-CoG-648328. H. D thanks the China Scholarship Council for financial support during her study at University College London.

# Supplementary Material for

# Quantum mechanical analysis of nonlinear optical response of interacting graphene nanoflakes

**1. Calculation of energy levels of graphene nanoflakes.**

In order to compute the polarizabilities of graphene nanoflakes (GNFs) we first calculate their electronic structures. Figure S1(a) shows the triangular and hexagonal structures of GNFs we studied in the main text. Our computational approach includes three-center terms in the Hamiltonian to take into account the core-electron interaction between distant atomic centers. The low energy optical response of GNF is dominated by excitations of the $\pi$ valence band, formed by electrons residing in the carbon $2p_z$ orbitals oriented perpendicularly with respect to the graphene plane, as per Fig. S1(b) and populated on average with one electron per carbon site.

In our model we used the $2p_z$ Clementi orbital[1] as a basis function on each atom, which takes the form of

$$\psi_{2p_z} = R_{2p}(r) Y_{2p_z}(\theta, \varphi), \tag{1}$$

where the radial part is

$$R_{2p}(r) = \frac{1}{\sqrt{6}} \frac{Zr}{na_0} Z^{\frac{3}{2}} e^{-\frac{Zr}{na_0}}, \tag{2}$$

and the angular part is

$$Y_{2p_z}(\theta, \varphi) = \sqrt{\frac{3}{4\pi}} \cos\theta = \sqrt{\frac{3}{4\pi}} \frac{z}{r}. \tag{3}$$

Here, $Z = 3.136$ is the effective nuclear charge for the $2p_z$ orbital of a carbon atom[1], $n$ is the orbital number ($n = 2$ in this case) and $a_0$ is the Bohr radius.

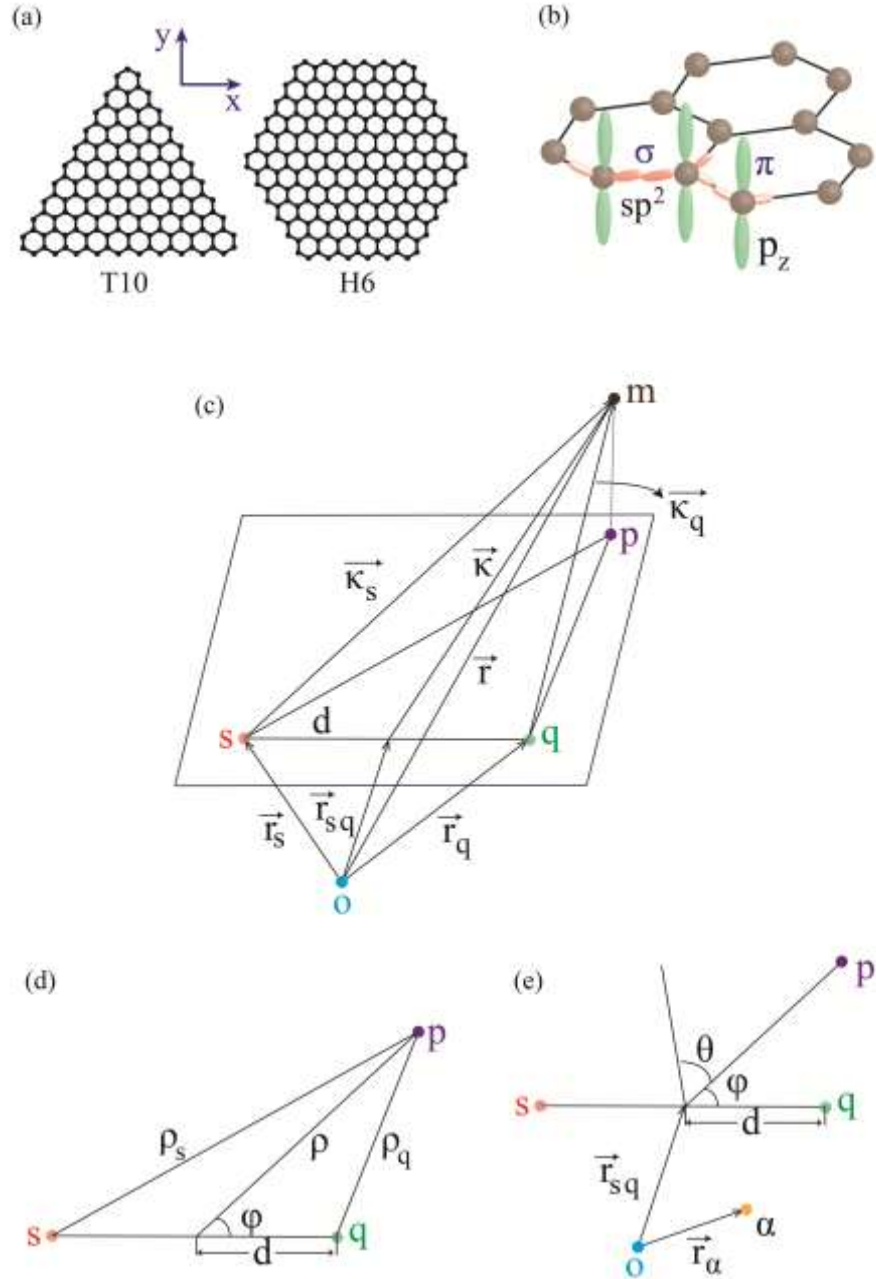

Fig. S1. (a) Schematics of triangular and hexagonal GNFs. T*n* and H*n* denote triangular and hexagonal GNFs, respectively, where *n* is the number of 6-atom hexagons on the side of GNFs. (b) $sp^2$ hybridization of carbon atoms of graphene. (c), (d), (e) Definitions of the vectors used in the calculations.

The Hamiltonian operator for a single electron in the GNF is

$$\hat{H} = -\frac{\hbar}{2m}\nabla_{\vec{r}}^2 - \sum_{\alpha=1}^{N}\frac{Z_{eff}e^2}{\vec{r}-\vec{r}_{0\alpha}}, \qquad (4)$$

where the value of effective core charge, $Z_{eff} = 0.637$, has been adjusted so that the computed HOMO-LUMO gap of the simplest GNF structure (i.e. benzene) is 6 eV[2] (see Fig. S2(a)).

As the Hamiltonian (S4) shows, in our treatment, each electron in the GNF is influenced by the core-potential of all the carbon atoms of the flake. Please note that in this treatment there is no nonlinear interaction between atoms or electrons, the nonlinear response of the system coming from the interaction with an external optical field (this is in contrast to the treatment in Ref.[3], where each atom is represented as a "dipole", and their interaction is of nonlinear nature).

The electronic structure is computed using the Schrödinger equation

$$\hat{H}\psi(\vec{r}) = E\psi(\vec{r}), \qquad (5)$$

where $E$ and $\psi(\vec{r})$ are the eigenenergy and eigenfunction, respectively. This eigenfunction is expanded in the atom centered basis functions as

$$\psi(\vec{r}) = \sum_{q=1}^{N} c_q \psi_q(\vec{r}) = \sum_{q=1}^{N} c_q \psi_{2p_z}(\vec{r}-\vec{r}_{oq}). \qquad (6)$$

Combining equations (5) and (6) yields

$$\sum_{q=1}^{N}\hat{H}c_q\psi_q = E\sum_{q=1}^{N}c_q\psi_q. \qquad (7)$$

We then multiply each side of this equation from the left by $\psi_s^*(\vec{r})$ and integrate over all space:

$$\sum_{q=1}^{N}c_q\int\psi_s^*\hat{H}\psi_q d\vec{r} = E\sum_{q=1}^{N}c_q\int\psi_s^*\psi_q d\vec{r}. \qquad (8)$$

Now we define the matrix elements

$$\hat{H}_{sq} = \int \psi_s^* \hat{H} \psi_q d\vec{r}, \tag{9a}$$

$$\hat{S}_{sq} = \int \psi_s^* \psi_q d\vec{r}, \tag{9b}$$

so that the Schrödinger equation is truncated to a finite generalized eigenvalue problem with eigenvectors, $\hat{c}$, and eigenvalues, $E$

$$\hat{H}\hat{c} = E\hat{S}\hat{c}, \tag{10}$$

where $\hat{c} = (c_1, c_2, ..., c_N)^T$ is an $N$-dimensional column vector.

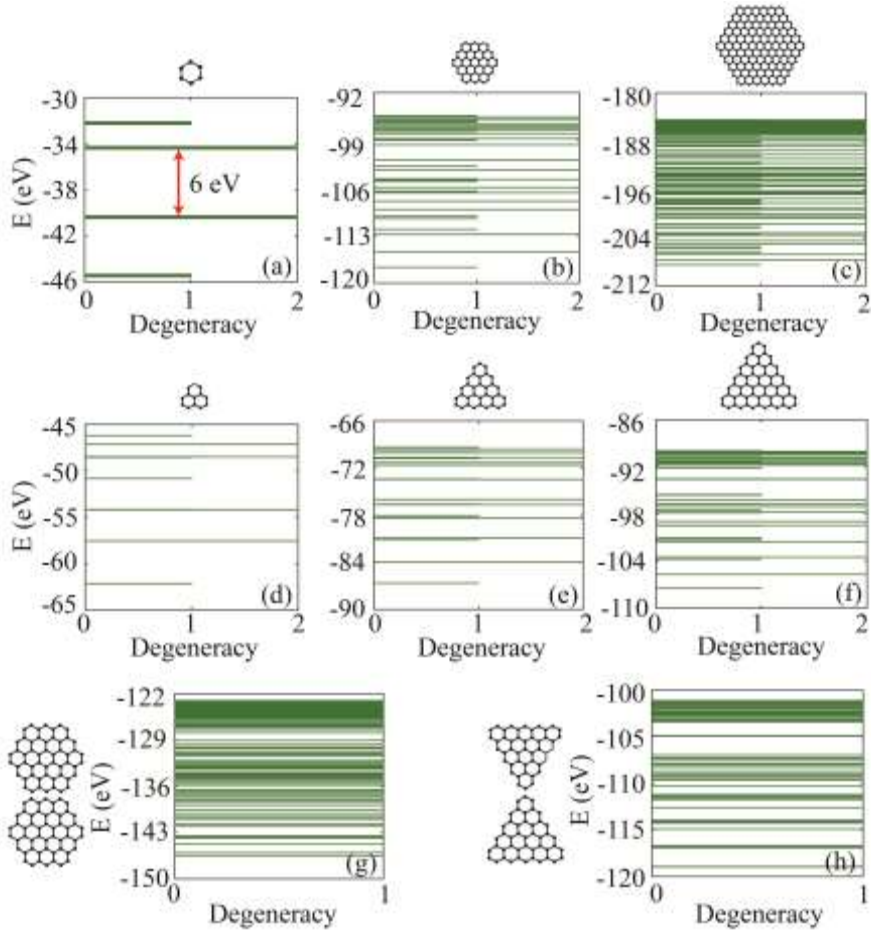

Fig. S2. (a), (b), (c) Energy levels of hexagonal GNFs with different sizes. (d), (e), (f) The same as in the top panels, but calculated for triangular GNFs. (g), (h) Energy levels of dimers made of hexagonal and triangular GNFs, respectively, with separation distance $D = 1a$, where $a = 1.42$ Å is the lattice constant. The red arrow in (a) indicates the HOMO-LUMO gap for single benzene.

The matrix element $\hat{H}_{sq}$ is written as the sum of two-center and three-center terms

$$\hat{H}_{sq} = E_0 \hat{S}_{sq} - \hat{I}_{sq} - \hat{I}_{sq}^{\alpha}, \qquad (11)$$

where $E_0 = \dfrac{Z_{eff}\varepsilon_0}{n^2}$, and $\varepsilon_0 = -13.6\,\text{eV}$.

$$\hat{I}_{sq} = \frac{Z_{eff}e^2}{2}\int \left(\frac{1}{|\vec{r}-\vec{r}_s|} + \frac{1}{|\vec{r}-\vec{r}_q|}\right)\psi_s^*\psi_q d\vec{r}, \qquad (12)$$

and

$$\hat{I}_{sq}^{\alpha} = \sum_{\alpha \neq s,q} Z_{eff}e^2 \int \frac{\psi_s^*\psi_q}{|\vec{r}-\vec{r}_\alpha|} d\vec{r}. \qquad (13)$$

Here $\vec{r}_s, \vec{r}_q$ and $\vec{r}_\alpha$ are the position vectors of the carbon atoms labeled by $s$, $q$, and $\alpha$, respectively.

By solving equation (10), one obtains the energy levels of a GNF. The energy levels of the hexagonal and triangular GNFs, calculated for GNFs with different sizes, are shown in Figs. S2(a)- S2(c) and Figs. S2(d)- S2(f), respectively. We also show in Figs. S2(g) and S2(h) the energy levels of hexagonal and triangular GNF dimers, respectively, separated by $D = 1a$, where $a = 1.42$ Å is the C-C bond length.

In what follows, we describe the approach used to evaluate the multi-center integrals. We convert the spherical coordinates to a cylindrical system, as per Figs. S1(c)-S1(e), where the lengths of vectors $\vec{\kappa}_s$ and $\vec{\kappa}_q$ are expressed as

$$|\vec{\kappa}_s| = \sqrt{\rho^2 + d^2 + 2\rho d \cos(\varphi) + z^2}, \qquad (14a)$$

$$|\vec{\kappa}_q| = \sqrt{\rho^2 + d^2 - 2\rho d \cos(\varphi) + z^2}, \qquad (14b)$$

where $\rho$, $\varphi$, and $z$ are the cylindrical coordinates and $d$ is the length of vector **d**, namely it is half of the distance between the carbon atoms $s$ and $q$. Therefore, the basis functions centered at $s$ and $q$ are written as:

$$\psi_s = Aze^{-\frac{Z}{na_0}|\vec{\kappa}_s|}, \qquad (15a)$$

$$\psi_q = Aze^{-\frac{Z}{na_0}|\vec{\kappa}_q|}, \tag{15b}$$

where $A = \frac{1}{\sqrt{8\pi}} \frac{Z}{na_0} (\frac{Z}{a_0})^{\frac{3}{2}}$ is a normalization constant. Therefore, $\hat{S}_{sq}$, $\hat{I}_{sq}^{\alpha}$, and $\hat{I}_{sq}$ are expressed as follows:

$$\hat{S}_{sq} = \int A^2 z^2 e^{-\frac{Z}{na_0}(|\hat{\kappa}_s|+|\hat{\kappa}_q|)} \rho d\rho d\varphi d\theta, \tag{16}$$

$$\hat{I}_{sq}^{\alpha} = \sum_{\alpha \neq s,q} \frac{A^2 z^2 e^{-\frac{Z}{na_0}(|\hat{\kappa}_s|+|\hat{\kappa}_q|)}}{\sqrt{\rho^2 + |\vec{R}_{sq}^{\alpha}|^2 - 2\rho|\vec{R}_{sq}^{\alpha}|\cos\theta + z^2}} \rho d\rho d\varphi d\theta, \tag{17}$$

and

$$\hat{I}_{sq} = \int (\frac{1}{|\vec{\kappa}_s|} + \frac{1}{|\vec{\kappa}_q|}) A^2 z^2 e^{-\frac{Z}{na_0}(|\vec{\kappa}_s|+|\vec{\kappa}_q|)} \rho d\rho d\varphi d\theta, \tag{18}$$

where $|\vec{R}_{sq}^{\alpha}|$ is the length of vector $\vec{R}_{sq}^{\alpha} = \vec{r}_{sq} - \vec{r}_{\alpha}$ and $\theta$ is the angle between the vector $\vec{r}$ and the plane of the graphene lattice, The integrals (16), (17), and (18) have been evaluated numerically.

**II. Quantum perturbative approach to the linear and nonlinear optical response of graphene nanoflakes.**

We model the optical response of GNFs using a well-known quantum perturbative approach[4]. The linear polarizability, $\alpha_{ij}(\omega)$, and the nonlinear polarizabilities corresponding to second-harmonic generation, $\beta_{ijk}(2\omega)$, and third-harmonic generation, $\gamma_{ijk}(3\omega)$ are given by

$$\alpha_{ij}(\omega) = \sum_{g,m}^{transition} (\frac{\mu_{gm}^i \mu_{mg}^j}{\omega_{mg} - \omega} + \frac{\mu_{gm}^j \mu_{mg}^i}{\omega_{mg}^* - \omega}), \tag{19}$$

$$\beta_{ijk}(2\omega) = P_l \sum_{g,m,n}^{transition} [\frac{\mu_{gn}^i \mu_{nm}^j \mu_{mg}^k}{(\omega_{ng}-2\omega)(\omega_{mg}-\omega)} + \frac{\mu_{gn}^j \mu_{nm}^i \mu_{mg}^k}{(\omega_{ng}^*+\omega)(\omega_{mg}-\omega)}$$
$$+ \frac{\mu_{gn}^j \mu_{nm}^k \mu_{mg}^i}{(\omega_{ng}^*+\omega)(\omega_{mg}^*+2\omega)}]$$
(20)

$$\gamma_{kjih}(3\omega) = P_l \sum_{g,m,n,v}^{transition} [\frac{\mu_{gv}^k \mu_{vn}^j \mu_{nm}^i \mu_{mg}^h}{(\omega_{vg}-3\omega)(\omega_{ng}-2\omega)(\omega_{mg}-\omega)} + \frac{\mu_{gv}^j \mu_{vn}^k \mu_{nm}^i \mu_{mg}^h}{(\omega_{vg}^*+\omega)(\omega_{ng}-2\omega)(\omega_{mg}-\omega)}$$
$$+ \frac{\mu_{gv}^j \mu_{vn}^i \mu_{nm}^k \mu_{mg}^h}{(\omega_{vg}^*+\omega)(\omega_{ng}^*+2\omega)(\omega_{mg}-\omega)} + \frac{\mu_{gv}^j \mu_{vn}^i \mu_{nm}^h \mu_{mg}^k}{(\omega_{vg}^*+\omega)(\omega_{ng}^*+2\omega)(\omega_{mg}^*+3\omega)}]$$

(21)

where $\vec{\mu} = e\vec{r}$ is the dipole moment operator, $\mu_{gm} = \int \psi_g^* \vec{\mu} \psi_m d\vec{r}$, is the transition dipole moment, $\omega_{mg} = \frac{E_m - E_g}{\hbar} - i\eta$, $|\psi_m\rangle = \sum_l c_{ml} |\psi_l\rangle$, $E_m$ and $c_m = \sum_l c_{ml}$ are the eigenenergy and engenvector of the eigenstae $m$, respectively. Moreover, $g$, $m$, as well as $n$, $v$ are labels used to distinguish between levels of transitions and $\eta = 0.1$eV is related to the lifetime of excited states. Here, we have introduced the intrinsic permutation operator, $P_l$, defined such that the expression that follows it is to be summed over all permutations of the Cartesian indices: $i$, $j$, $k$, and $h$.

### III. Linear and nonlinear polarizability units.

Throughout the paper we used atomic units for the linear and nonlinear polarizabilities $\alpha$, $\beta$, and $\gamma$, $eV$ for energy, and $nm$ for length. As Système International (SI) units are widely used, we present the conversion factors between SI and atomic units.

In equations (19), (20), and (21), the transition dipole moment elements are in a.u., where 1 a.u. of the dipole is 1 electron charge times the Bohr radius. Specifically, we have

$$ea_0 = 1.602 \times 10^{-19} \times 0.529 \times 10^{-10} (Cm).$$
(22)

Moreover, in equations (19), (20), and (21) we use $eV$ as the energy unit, $\omega$ must be multiplied by $\hbar$, therefore,

$$\frac{1}{\hbar\omega} = \frac{1}{1.602\times 10^{-19}} (J^{-1}). \tag{23}$$

All the coefficients and units in equations (19), (20), and (21) must be multiplied together. Thus, the conversion factors from the calculated value to the SI unit for $\alpha$, $\beta$, and $\gamma$ are given by

$$[\alpha] = \frac{(ea_0)^2}{\hbar\omega} = \frac{(1.602\times 10^{-19} \times 0.529\times 10^{-10})^2}{1.602\times 10^{-19}} C^2 m^2 J^{-1}$$
$$= 4.48\times 10^{-40} C^2 m^2 J^{-1}, \tag{24}$$

$$[\beta] = \frac{(ea_0)^3}{(\hbar\omega)^2} = \frac{(1.602\times 10^{-19} \times 0.529\times 10^{-10})^3}{(1.602\times 10^{-19})^2} C^3 m^3 J^{-2}$$
$$= 2.37\times 10^{-50} C^3 m^3 J^{-2}, \tag{25}$$

and

$$[\gamma] = \frac{(ea_0)^4}{(\hbar\omega)^3} = \frac{(1.602\times 10^{-19} \times 0.529\times 10^{-10})^4}{(1.602\times 10^{-19})^3} C^4 m^4 J^{-3}$$
$$= 1.25\times 10^{-60} C^4 m^4 J^{-3}. \tag{26}$$